\documentclass{PoS}

\usepackage{amsfonts,amssymb,amsmath,bm}
\usepackage{graphicx}

\newcommand{\be}{\begin{equation}}
\newcommand{\bea}{\begin{eqnarray}}
\newcommand{\ee}{\end{equation}}
\newcommand{\eea}{\end{eqnarray}}

\def\s#1{{\scriptscriptstyle #1}}
\def\gb{\bm{\Gamma}}

\title{On the dynamics of the Kugo-Ojima function}

\ShortTitle{On the dynamics of the Kugo-Ojima function}

\author{\speaker{Daniele Binosi}\thanks{Talk based on work done in collaboration with Arlene C. Aguilar 
and Joannis Papavassiliou}\\
       ECT* European Centre for Theoretical Studies in Nuclear Physics and Related Areas\\
        E-mail: \email{binosi@ect.it}}

\abstract{
In this talk, after reviewing the dynamical gluon mass generation mechanism within the pinch technique framework and its phenomenological predictions, we will introduce the modern formulation of the pinch technique which makes extensive use of the Batalin-Vilkovisky quantization formalism. In this framework a certain auxiliary function $\Lambda_{\mu\nu}(q)$ -- and its associated form factors $G(q^2)$ and $L(q^2)$ -- play a prominent role. After showing that in the (background) Landau gauge $\Lambda_{\mu\nu}(q)$ fully constrains the QCD ghost sector, we show that $G(q^2)$ coincides with the Kugo-Ojima function $u(q^2)$, whose infrared behavior has traditionally served as the standard criterion for the realization of 
the Kugo-Ojima confinement mechanism. The determination of the behavior of $G(q^2)$ (and therefore of the Kugo-Ojima function) for all momenta through a combination of the available lattice data on the gluon and ghost propagators as well as the dynamical equation $G(q^2)$ satisfies, will be then discussed. In particular we will show that ({\it i}) in the deep infrared the 
function deviates considerably from the value associated with the realization of the Kugo-Ojima 
confinement scenario, and ({\it ii}) establish the dependence on the renormalization point of $u(q^2 )$, and especially of its value at $q^2=0$.}

\FullConference{International Workshop on QCD Green's Functions, Confinement, and Phenomenology - QCD-TNT09\\
		 September 07 - 11 2009\\
		 ECT Trento, Italy}

\begin{document}

In the last two years, ab-initio lattice gauge theory computations using extremely large volumes have firmly established that (in the Landau gauge) the QCD gluon propagator and the ghost dressing function are infrared (IR) finite and non-vanishing~\cite{Cucchieri:2007md,Bogolubsky:2007ud}.  Specifically choosing an $R_\xi$ type of gauge and defining the gluon propagator cofactor $\Delta$, and the ghost dressing function $F$ as
\be
\Delta_{\mu\nu}(q)=-\mathrm{i}\left[P_{\mu\nu}(q)\Delta(q^2)+\xi\frac{q_\mu q_\nu}{q^2}\right], \qquad D(q^2)=\mathrm{i}\frac{F(q^2)}{q^2},
\ee
where $P_{\mu\nu}(q)=g_{\mu\nu}-q_\mu q_\nu / q^2$ is the transverse projector, $\Delta^{-1}(q^2)=q^2+\mathrm{i}\Pi(q^2)$ [with $\Pi_{\mu\nu}(q^2)=P_{\mu\nu}(q)\Pi(q^2)$ the gluon self-energy], and $D(q^2)$ is the ghost propagator, lattice results tells us  that (Euclidean space)
\be
\Delta^{-1}(0)>0,\qquad \mathrm{and}\qquad F(0)>0.
\ee
The issue of explaining these clean lattice results from the point of view of the continuum formulation of the theory has therefore become an increasingly interesting topic, for obtaining them is bound to expose a QCD fundamental dynamical mechanism at work. 

Indeed, such a mechanism is provided by the dynamical generation of a gluon mass~\cite{Cornwall:1982zr} through the non-perturbative realization of the well-known mechanism described long ago by Schwinger~\cite{Schwinger:1951ex}. Schwinger mechanism shows that if for some reason the dimensionless vacuum polarization $\Pi(q^2)/q^2$ behaves as a simple pole with positive residue $\mu^2$ at $q^2=0$ (an there is no physical principle precluding this possibility) then $\Delta^{-1}(q^2)=q^2+\mu^2$: Thus the vector meson (which is massless in the absence of interactions) becomes massive, with $\Delta^{-1}(0)=\mu^2$. When the theory is strongly coupled, as it happens with QCD in the IR, strong binding may generate zero-mass bound state excitations which, notwithstanding the fact that they do not generate from the spontaneous breakdown of any symmetry, acts like  massless, composite and longitudinally coupled (dynamical) Nambu-Goldstone bosons~\cite{Jackiw:1973tr}. 

\begin{figure}[!t]
\includegraphics[scale=2.0]{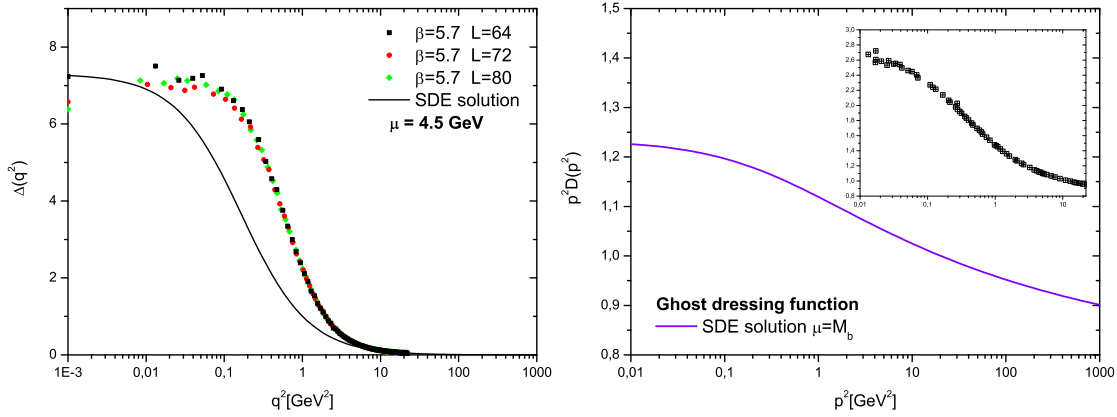}
\caption{\label{fig1}{\it Left Panel}: The numerical solution for the gluon propagator from the PT modified SDE
(black solid line) compared to the 
lattice data of~\cite{Bogolubsky:2007ud}. {\it Right panel}: The ghost dressing function $p^2D(p^2)$ 
obtained from the SDE. In the inset we show  
the lattice data for the same quantity; notice the absence of any enhancement in both cases.}
\end{figure}

The implementation of this mechanism within the pinch technique (PT) framework~\cite{Cornwall:1982zr,Cornwall:1989gv,Binosi:2002ft}, gives rise to two complementary effects, which appears at the level of the QCD Schwinger-Dyson equations (SDEs), and of the effective low-energy theory which describes QCD in the IR sector, eventually providing a confinement mechanism.
\begin{itemize}

\item{\it Schwinger-Dyson equations.} The systematic exploitation of the underlying BRST symmetry provided by the PT originate drastic modifications to the Green's functions of the theory and the corresponding SDEs which describes their dynamics; in particular the new SDE obtained for the gluon propagator lend itself to a novel truncation scheme that respects gauge invariance at every level of the dressed loop expansion~\cite{Binosi:2007pi}. Once this scheme is used together with the assumption that the three-gluon vertex contains dynamical massless poles ($\sim 1/q^2$) triggering the Schwinger-mechanism, one obtains IR finite solutions for the propagator~\cite{Aguilar:2008xm}. These solutions are of the type $\Delta^{-1}(q^2)=q^2+m^2(q^2)$ with the mass depending non-trivially on the momentum transfer, and with (obviously) $\Delta^{-1}(0)=m^2(0)=m^2_0>0$ (Fig.~\ref{fig1}, left panel). In addition of  taming the IR divergences intrinsic to perturbation theory (Landau pole) this mass forces $F(q^2)$ to stay IR finite with $F(0)>0$ (Fig.~\ref{fig1}, right panel), and no enhancement is found~\cite{Boucaud:2008ky}.
 
\item{\it Effective low-energy theory.} An effective low-energy field 
theory for describing the gluon mass   
is  the  gauged non-linear sigma model  known  as ``massive
gauge-invariant Yang-\linebreak Mills''~\cite{Cornwall:1979hz}, with    
Lagrangian density  
\begin{equation}
{\cal L}_{\s{\mathrm{MYM}}}= \frac{1}{2} F_{\mu\nu}^2 - 
m_0^2 {\rm Tr} \left[A_{\mu} - {g}^{-1} U(\theta)\partial_{\mu} U^{-1}(\theta) \right]^2,
\label{nlsm}
\end{equation}
where 
$A_{\mu}= \frac{1}{2\mathrm{i}}\sum_{a} \lambda_a A^{a}_{\mu}$, the $\lambda_a$ are the SU(3) generators
(with  ${\rm Tr} \lambda_a  \lambda_b=2\delta_{ab}$), 
and the $N\times N$
unitary matrix $U(\theta) = \exp\left[\mathrm{i}\frac{1}{2}\lambda_a\theta^{a}\right]$ 
describes the scalar fields $\theta_a$.  
Note that ${\cal L}_{\s{\mathrm{MYM}}}$ is locally gauge-invariant under the combined gauge transformation 
\be
A^{\prime}_{\mu} = V A_{\mu} V^{-1} - {g}^{-1} \left[\partial_{\mu}V \right]V^{-1}, 
\qquad
U^\prime = U(\theta^\prime) = V U(\theta),
\label{gtransfb}
\ee
for any group matrix $V= \exp\left[i\frac{1}{2}\lambda_a\omega^{a}(x)\right]$, where 
$\omega^{a}(x)$ are the group parameters. 
One might think that, by employing (\ref{gtransfb}), the fields  $\theta_a$ can always 
be transformed to zero, but this is not so if the $\theta_a$ contain vortices.
To use the ${\cal L}_{\s{\mathrm{MYM}}}$ in (\ref{nlsm}), one solves the equations of motion for $U$ 
in terms of the gauge potentials and substitutes the result in the equations for the gauge potential.  
One then finds the Goldstone-like massless modes mentioned above.
This model admits  vortex
solutions~\cite{Cornwall:1979hz},  with a  long-range pure  gauge term  in  their potentials,
which endows  them with a topological quantum  number corresponding to
the center  of the gauge group  [$Z_N$ for $SU(N)$], and  is, in turn,
responsible for quark  confinement and gluon screening. 
Specifically, center vortices of  thickness $\sim m_0^{-1}$,  form a condensate because their entropy
(per  unit  size) is  larger  than  their  action.  This  condensation
furnishes an  area law to  the fundamental representation  
Wilson loop, thus confining quarks~\cite{Cornwall:1982zr,Cornwall:1979hz}. 
In addition, the adjoint potential shows a
roughly linear  regime followed by string breaking  when the potential
energy is about $2m_0$,  corresponding to gluon screening~\cite{Bernard:1982my}.
\end{itemize}

Thus, summarizing, in this picture the non-perturbative QCD dynamics generate an effective, momentum dependent mass, without affecting the local $SU(3)$ invariance, which remains intact. This provides in turn very definite predictions about the IR behavior of the theory (all in agreement with lattice studies):  ({\it i}) the gluon propagator is IR finite; ({\it ii}) in the Landau gauge the ghost remains massless, but with a finite dressing function;  the presence of center vortex solutions providing ({\it iii}) an area law for the Wilson loop and ({\it iv})  a roughly linear behavior for the adjoint potential followed by string breaking.

A different set of predictions is obtained within the Kugo-Ojima (KO) scenario, which also establishes a highly non-trivial link
between confinement and the infrared behavior of some  fundamental Green's functions of QCD~\cite{Kugo:1979gm}. In the
KO confinement picture  one starts by observing that the equation of motion for the gauge field can be written in the 
Maxwell-like form $\partial^\nu F^a_{\mu\nu}+\{Q_{\s{\mathrm{BRST}}},({\cal D}_\mu c)^a\}=gJ^a_\mu$, with $J^a_\mu$ 
the Noether current of the global color symmetry, ${\cal D}_\mu$ the usual covariant derivative,  and $Q_{\s{\mathrm{BRST}}}$
the BRST charge operator. On the other hand, if one adds to a Noether current the derivative of an antisymmetric rank 2 tensor
(say $\partial^\mu f_{[\mu,\nu]}$) the resulting current is still preserved and the corresponding charge correctly generates
rotations in color space. Thus one is,  at least in principle, allowed to define the color charge $Q^a$ in the BRST exact form
\be
Q^a=\int\!\mathrm{d}^3x\,\left(J^a_0-\frac1g\partial^\nu F^a_{0\nu}\right)=\int\!\mathrm{d}^3x\,\frac1g\{Q_{\s{\mathrm{BRST}}},({\cal D}_0c)^a\}.
\label{c-charge}
\ee
Confinement is then a direct consequence of the above relation, since for any physical state specified by the condition $Q_{\s{\mathrm{BRST}}}\vert\mathrm{phys}\rangle=0$ one has $\langle\mathrm{phys}\vert Q^a\vert\mathrm{phys}'\rangle=0$, which implies that all physical states are color-singlets~\cite{Kugo:1979gm}.

The problem is however that the volume integral in Eq.~(\ref{c-charge}) does not converge, due to the presence of massless one particle contributions to $J^a_\mu$, $\partial^\nu F^a_{\mu\nu}$, and $\{Q_{\s{\mathrm{BRST}}},({ \cal D}_\mu c)^a\}$ (the so-called quartet mechanism). Without entering into any detail, one has that a solution to this problem is provided by introducing some suitable weights $v$, $w$ and $u$ respectively, so that a well defined charge is given
by
\be
Q^a=\int\!\mathrm{d}^3x\,\left(J^a_0+\frac vw\partial^\nu F^a_{0\nu}\right), \qquad gv=-w + (1+u)
\label{c-charge-wd}
\ee
where the relation represents the condition between the different weights for the cancellation of the aforementioned one-particle massless contributions. Thus, requiring that Eq.~(\ref{c-charge-wd}) coincides with the BRST exact expression (\ref{c-charge}) implies $v/w = -1/g$, and therefore the KO confinement criterion~\cite{Kugo:1979gm}
\be
1+u=0.
\label{KOs}
\ee

It turns out that in the Landau gauge the so-called KO parameter $u$ is linked to the IR limit $q^2\to0$ of a certain Green's function; more precisely one has
\bea
& & \int\!\mathrm{d}^4x\, \mathrm{e}^{-\mathrm{i}q\cdot(x-y)}\langle T\big[\left({\cal D}_\mu c\right)_x^m\left({\cal D}_\mu \bar c\right)_y^n\big]\rangle=
-\frac{q_\mu q_\nu}{q^2}\delta^{mn}+P_{\mu\nu}(q)\delta^{mn}u(q^2), \label{id-0}\\
& & \lim_{q^2\to0}u(q^2)=u(0)=u.
\label{id-1}
\eea
In addition, there is a powerful BRST identity relating the KO-function $u(q^2)$ and the ghost dressing function $F(q^2)$, namely~\cite{Kugo:1995km, Grassi:2004yq}
\be
F^{-1}(q^2)=1+u(q^2)+w(q^2),
\label{BRST-1}
\ee
where for now $w(q^2)$ is an unspecified function, which under very general conditions is such that $w(0)=0$. Then, Eq.~(\ref{BRST-1}) tells us that the KO confinement scenario predicts an IR divergent ghost dressing function, at odds with the aforementioned large volumes lattice simulations\footnote{The same prediction is obtained when implementing the (original) Gribov-Zwanziger (GZ) 
horizon condition~\cite{Gribov:1977wm}: in the IR region the ghost propagator diverges more rapidly than 
at tree-level. Furthermore, it has been also argued that the Landau gauge gluon propagator should vanish in the same limit. The prediction for the gluon propagator in the KO scenario is instead that it is 
less  divergent  than  the  tree-level  expression  which,  evidently, encompasses  the IR-finite  gluon propagator  as a  special  case, even though, up until recently, the focus had been placed rather on the ``vanishing'' solutions, given 
that they satisfy simultaneously both the KO and GZ requirements. }.

An issue that to the best of our knowledge has never been thoroughly addressed in the KO scenario is how renormalization affects the proof of the central identity~(\ref{KOs}). On the other hand, and at a less formal level, one should notice that once the relations~(\ref{id-1}) are proved, the KO function is on a par with any other QCD Green's function: quantum  corrections   will  set  in,   and  the  whole   procedure  of regularization and renormalization should  be applied. This in general
implies the unavoidable appearance of a dependence in the Green's  function~(\ref{id-1}) on a ``sliding  scale''  $\mu$. 
This  dependence disappears {\it only}  ({\it i}) when combining individual Green's functions to form observables, such 
as $S$ matrix elements, or ({\it ii}) when forming  very special (and very well studied) 
products of  Green's  functions  {\it and } the gauge coupling of the theory. The latter are the so-called renormalization-group (RG) invariant combinations, with the product $e^2 \Delta$ in QED constituting probably the most celebrated 
text-book case. Whether a product of Green's functions forms a RG-invariant combination or not is determined on  {\it formal } grounds, from the Ward-Takahashi or Slavnov-Taylor identities satisfied by the quantities involved 
({\it e.g.}, the famous $Z_1=Z_2$ of QED).  The Green's function of Eq.~(\ref{id-1}) is definitely not an RG-invariant~\cite{Aguilar:2009pp}, and therefore picks up a non-trivial dependence on  $\mu$. Thus, one has $u(q^2,\mu^2)$, and in particular, in the deep IR limit, $u=u(0,\mu^2)$.

Now, if the KO confinement criterion~(\ref{KOs}) were satisfied, the ghost dressing function must diverge as $q^2\to0$, 
due to the BRST identity (\ref{BRST-1}).  This will in turn make  the $\mu$ dependence of
the KO function (which is  however still there) irrelevant, since $u(q^2,\mu^2)$ will
be then driven  to -1  in the deep IR limit for any value of $\mu$. On the other hand, lattice simulations tells us that the ghost dressing function is finite, and it is interesting to determine explicitly the $\mu$-dependence of the KO function.

Quite remarkably this issue can be thoroughly studied within the modern formulation of the PT (by means of the Batalin-Vilkoviski quantization formalism~\cite{Batalin:1977pb}) exploiting in particular the PT correspondence~\cite{Binosi:2002ft,Denner:1994nn} with the background field method (BFM)~\cite{Abbott:1980hw}. Indeed, it turns out that~\cite{Grassi:2004yq,Aguilar:2009pp}, in the (background) Landau gauge, the Kugo-Ojima function coincides with the form factor $G(q^2)$ multiplying $g_{\mu\nu}$ in the Lorentz decomposition of a certain auxiliary function $\Lambda_{\mu\nu}(q)$ which enters in all the so-called ``background-quantum'' identities~\cite{Grassi:1999tp}, {\it i.e.}, the infinite tower of non-trivial relations connecting the BFM Green's functions to the conventional ones ({\it e.g.}, calculated in the $R_\xi$ gauges).  In addition, $G(q^2)$ is a key element in the aforementioned new SDEs  that can be truncated in  a manifestly gauge invariant way. 

Let us recall that in the BV formulation of Yang-Mills theories~\cite{Batalin:1977pb}, 
one starts by introducing certain sources (called anti-fields in what follows) 
that describe the renormalization of composite operators; the latter class 
of operator is in fact bound to appear in such theories due to the non-linearity 
of the BRST transformation of the elementary fields. In much the same way, 
the quantization of the theory in a background field type of gauge requires, in addition 
to the aforementioned anti-fields, the introduction of new sources which couple to 
the BRST variation of the background fields~\cite{Grassi:1999tp}. 
These sources are sufficient for implementing the full set of symmetries of a 
non-Abelian theory at the quantum level, and in the case of quarkless $SU(N)$ QCD, lead to the master equation
\be
\int\!\mathrm{d}^4x\left[\frac{\delta\Gamma}{\delta A^{*m}_\mu}\frac{\delta\Gamma}{\delta A^{m}_\mu}+\frac{\delta\Gamma}{\delta c^m}\frac{\delta\Gamma}{\delta\bar c^m}+B^m\frac{\delta\Gamma}{\delta\bar c^m}+\Omega^m_\mu\left(\frac{\delta\Gamma}{\delta \widehat{A}^m_\mu}-\frac{\delta\Gamma}{\delta A^m_\mu}
\right)\right]=0.
\label{me}
\ee
In the formula above, $\Gamma$ is the effective action, $A^*$ and $c^*$ the gluon and ghost anti-fields, $\widehat{A}$ 
is the gluon background field, and $\Omega$ the corresponding background source; finally $B$ denotes 
the Nakanishi-Lautrup multiplier for the gauge fixing condition.

To determine the complete algebraic structure of the theory we need two additional equations. 
The first one is the Faddeev-Popov equation, that controls the result of the contraction 
of an anti-field  leg with the corresponding momenta. In position space, it reads
\be
\frac{\delta\Gamma}{\delta \bar c^m}+\left(\widehat{\cal D}^\mu\frac{\delta\Gamma}{\delta A^*_\mu}\right)^m-\left({\cal D}^\mu\Omega_\mu\right)^m=0,
\label{FPE}
\ee
where
$({\cal D}^\mu \Phi)^m=\partial^\mu \Phi^m+gf^{mnr}A^n_\mu \Phi^r$ [in the case of $(\widehat{\cal D}^\mu \Phi)^m$ replace the gluon field $A$ with a background gluon field $\widehat{A}$].
The second one is the anti-ghost equation formulated in  the background field Landau gauge, which 
reads~\cite{Grassi:2004yq}
\be
\frac{\delta\Gamma}{\delta c^m}-\left(\widehat{\cal D}^\mu \frac{\delta \Gamma}{\delta \Omega_\mu}\right)^m-\left({\cal D}^\mu A^*_\mu\right)^m-f^{mnr}c^{*n}c^r
%-\mathrm{i}g\bar\psi^*t^a\psi+\mathrm{i}g\bar\psi t^a\psi^*
+f^{mnr}\frac{\delta\Gamma}{\delta B^n}\bar{c}^r=0,
\label{AGE}
\ee
This equation fully constrains the dynamics of the ghost field $c$, and implies that the latter 
will not get an independent renormalization constant.  The local form of the anti-ghost equation (\ref{AGE}) 
is only valid when choosing the background Landau gauge condition $(\widehat{\cal D}^\mu A_\mu)^m=0$; 
in the usual Landau gauge, $\partial^\mu A_\mu^m=0$, an integrated version of this equation is available. 
In fact, even though the results that follow will be derived for convenience in the background Landau gauge, they are valid also in the conventional Landau gauge of the $R_\xi$.

\begin{figure}[!t]
\begin{center}
\includegraphics[width=8cm]{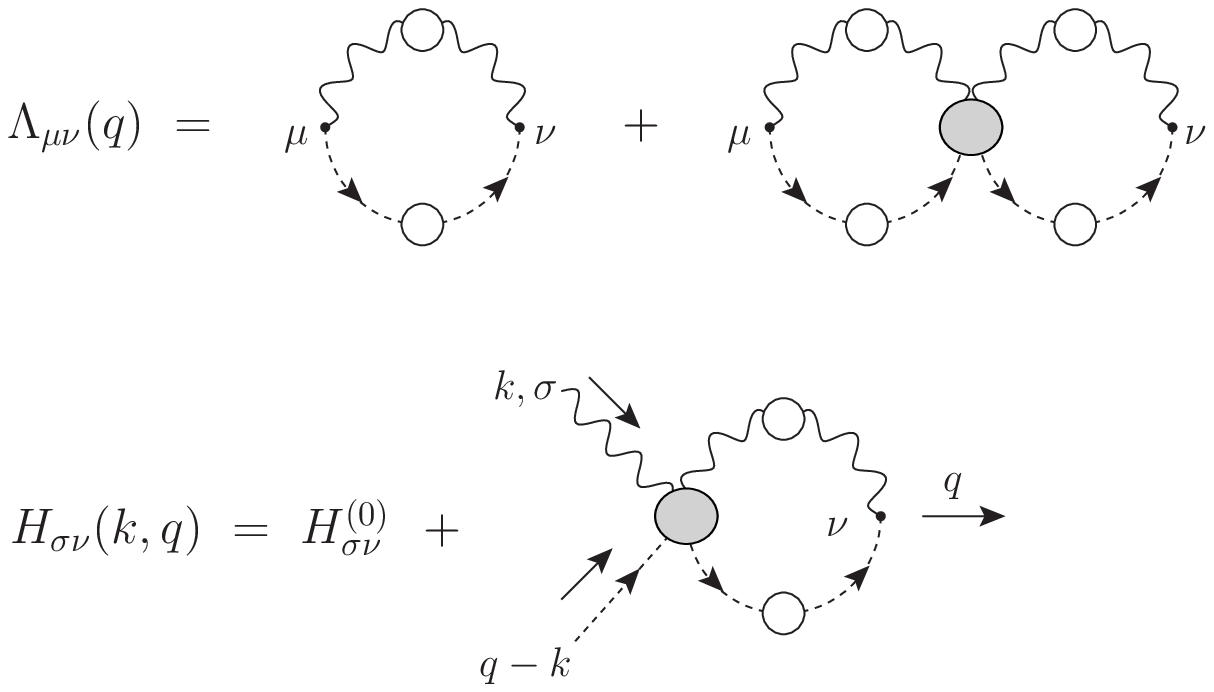}
\end{center}
\caption{Diagrammatic representation of the functions $\Lambda$ and $H$.}
\label{fig:Lambda-H}
\end{figure}

Now, differentiation of the functional (\ref{me}) with respect to a combination of fields containing 
at least one ghost field or two ghost fields and one anti-field (and setting the fields and sources 
to zero afterwards) will provide the Slavnov-Taylor identities of the theory. Differentiation 
with respect to a background source and background or quantum fields will provide, instead, the 
aforementioned background-quantum identities. 
Finally, differentiation of (\ref{FPE}) and (\ref{AGE}) with respect to fields and anti-fields or 
background sources give rise to relation among the different auxiliary ghost functions appearing in the theory.

The important point is that, when carrying out these differentiations, the following function appears (Fig.~\ref{fig:Lambda-H})
\bea
\mathrm{i}\Lambda_{\mu \nu}(q) &=&\Gamma_{\Omega_\mu A^*_\nu}(q) 
\ =\ g^2C_A
\int_k H^{(0)}_{\mu\rho}
D(k+q)\Delta^{\rho\sigma}(k)\, H_{\sigma\nu}(k,q),
\nonumber \\
&=&\mathrm{i} g_{\mu\nu} G(q^2) + \mathrm{i}\frac{q_{\mu}q_{\nu}}{q^2} L(q^2),
\label{LDec}
\eea
In the equations above, the color factor $\delta^{mn}$ has been factored out (as always in what follows), 
$C_{\rm {A}}$ represents the Casimir eigenvalue of the adjoint representation
[$C_{\rm {A}}=N$ for $SU(N)$], 
and \mbox{$\int_{k}\equiv\mu^{2\varepsilon}(2\pi)^{-d}\int\!d^d k$}, 
with $d=4-\epsilon$ the dimension of space-time. Finally, the function $H_{\mu\nu}(k,q)$ (see  Fig.~\ref{fig:Lambda-H} again)
is in fact a familiar object, for
it appears in the all-order Slavnov-Taylor identity
satisfied by the standard  three-gluon vertex;
it is also related to the full gluon-ghost vertex $\gb_{\mu}(k,q)$ by the identity
$q^\nu H_{\mu\nu}(k,q)=-\mathrm{i}\gb_{\mu}(k,q)$ [at tree-level, $H_{\mu\nu}^{(0)} = \mathrm{i}g_{\mu\nu}$ and $\gb^{(0)}_{\mu}(k,q)=-q_\mu$]. Indeed one finds the following results

\begin{figure}[!t]
\begin{center}
\includegraphics[width=8cm]{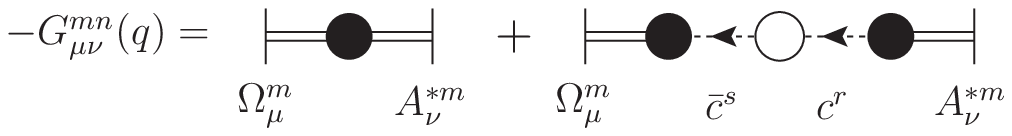}
\end{center}
\caption{Connected components contributing to the function $G^{mn}_{\mu\nu}(q)$.}
\label{connect}
\end{figure}

\begin{itemize}

\item[({\it i})] When differentiating the functional~(\ref{me}) with respect to a background source and a background gluon, on the 
one hand, and a background source and a quantum gluon, on the other, we can combine the resulting equation and trade the resulting two point functions for the corresponding propagators to get the important background-quantum identity
\be
\widehat{\Delta}^{-1}=\left[1+G(q^2)\right]^2\Delta^{-1}(q^2).
\label{gBQI}
\ee
The quantity $\widehat{\Delta}(q^2 )$ appearing on the left-hand side of the above equation captures the 
running of the QCD $\beta$ function, exactly as it happens with the QED vacuum polarization; 
for every value of the (quantum) gauge-fixing parameter one has (at the one loop level) $\Delta^{-1}(q^2)=q^2[1+bg^2\log(q^2/\mu^2)]$ where $b=11C_A/48\pi^2$. It is the identity (\ref{gBQI}) that plays a central role in the derivation of the new set of SDEs that can be truncated in manifestly gauge invariant way.

\item[({\it ii})] If we consider the background Landau gauge, differentiating the ghost equation (\ref{FPE}) with respect to a ghost field and a background source,  and the anti-ghost equation~(\ref{AGE}) with respect to a gluon anti-field and an anti-ghost,
we get the relations
\bea
\Gamma_{c\bar c}(q)=-\mathrm{i}q^\nu\Gamma_{c A^{*}_\nu}(q) &\qquad&
\Gamma_{\bar c\Omega_\mu}(q)=q_\mu+q^\nu\Lambda_{\mu\nu}(q),
\label{FPE_ghost} \\
\Gamma_{c A^{*}_\nu}(q)=q_\nu+q^\mu\Lambda_{\mu\nu}(q), &\qquad&
\Gamma_{c\bar c}(q)=-\mathrm{i}q^\mu\Gamma_{\bar c\Omega_\mu}(q).
\label{AGE-1}
\eea
Next,  contracting the first equation in~(\ref{AGE-1}) with $q^\nu$, and making use of  the first equation in~(\ref{FPE_ghost}), 
we see that the dynamics of the ghost sector is entirely captured by $\Lambda_{\mu\nu}(q)$, for one has
$\mathrm{i}\Gamma_{c\bar c}(q)=q^2+q^\mu q^\nu\Lambda_{\mu\nu}(q)$. Then,
introducing  the Lorentz decompositions
$\Gamma_{cA^*_\mu}(q)=q_\mu C(q^2)$ and $\Gamma_{\bar c \Omega_\mu}(q)=q_\mu E(q^2)$
we find the identities~\cite{Grassi:2004yq}
\bea
C(q^2) = E(q^2) = F^{-1}(q^2), \qquad
F^{-1}(q^2)&=&1+G(q^2)+L(q^2).
\label{ids}
\eea 
Recalling that the dimension of the gluon anti-field $A^*$ is three, 
while the dimension of the $\Omega$ source is one, power counting  shows that  
all functions appearing in Eqs.~(\ref{FPE_ghost}) and~(\ref{AGE-1}) are divergent, 
and in particular that the divergent part of $\Lambda_{\mu\nu}(q)$ can be 
proportional to $g_{\mu\nu}$ only~\cite{Grassi:2004yq,Aguilar:2009nf}, so that $L(q^2)$ is ultraviolet finite.

\item[({\it iii})] In the background Landau gauge the function appearing on the lhs of Eq.~(\ref{id-0}) is precisely given by
\be
-{\cal G}^{mn}_{\mu\nu}(q)=\frac{\delta^2 W}{\delta \Omega^{m}_\mu\delta A^{*n}_\nu},
\ee
where $W$ is the generator of the connected Green's functions, and the two connected diagrams contributing to  ${\cal G}_{\mu\nu}$ are shown in Fig.~\ref{connect}. Factoring out the color structure and making use of the identities~(\ref{ids}) one has
\bea
-\mathrm{i}{\cal G}_{\mu\nu}(q)&=&%\Gamma_{\Omega_\mu A^*_\nu}(q) 
\Lambda_{\mu\nu}(q)+ \Gamma_{\Omega_\mu\bar c}(q)D(q^2)\Gamma_{A^*_\nu c}(q)%\nonumber \\
=-\frac{q_\mu q_\nu}{q^2}+P_{\mu\nu}(q)G(q^2).
\label{id-2}
\eea
Passing to the Euclidean formulation, and comparing with Eq.~(\ref{id-0}), we then arrive at the important equality  
\be
u(q^2)=G(q^2).
\label{KO-G}
\ee
Then, the usual KO confinement criterion may be equivalently cast in the form: $1+G(0)=0$; moreover we see that the unspecified function $w(q^2)$ appearing in Eq.~(\ref{BRST-1}) coincides in fact with the $L(q^2)$ form factor appearing in ~(\ref{LDec}).

\end{itemize}

We thus see that in the (background) Landau gauge the single function $G(q^2)$ is an extremely interesting object to study, since its IR behavior determines much of the IR behavior of the theory. Now, keeping the vertices $\gb_\mu(k,q)$ and $H_{\mu\nu}(k,q)$ at their tree-level values, one finds that the dynamical equations for $F(q^2)$, $G(q^2)=u(q^2)$ and $L(q^2)$ are completely determined by the gluon and ghost propagator~\cite{Aguilar:2009nf}
\bea
F^{-1}(q^2)&=&1+g^2 C_{\rm {A}}\int_k \left[1- \frac{(k \cdot q)^2}{k^2 q^2}\right]\Delta (k)  D(k+q),\nonumber \\
u(q^2)& =& \frac{g^2 C_{\rm {A}}}{3}\int_k \left[2+ \frac{(k \cdot q)^2}{k^2 q^2}\right]\Delta (k)  D(k+q),\nonumber\\
L(q^2) &=& \frac{g^2 C_{\rm {A}}}{3}\int_k \left[1 - 4 \, \frac{(k \cdot q)^2}{k^2 q^2}\right]\Delta (k)  D(k+q).
\label{simple}
\eea

\begin{figure}[!t]
\begin{minipage}[b]{0.5\linewidth}
\centering
\includegraphics[scale=0.75]{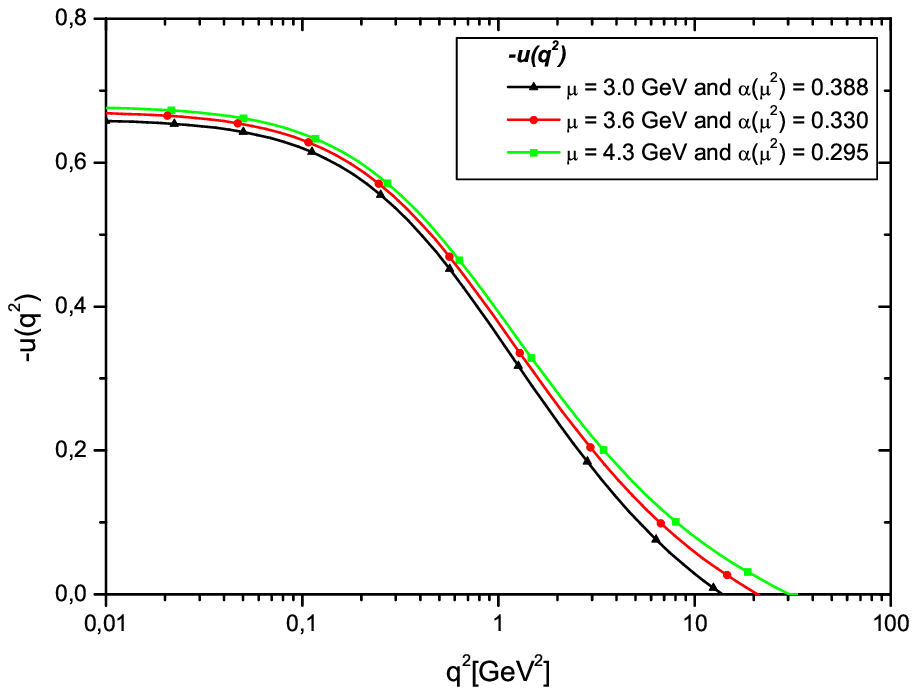}
\end{minipage}
%\hspace{0.5cm}
\begin{minipage}[b]{0.45\linewidth}
\centering
\includegraphics[scale=0.75]{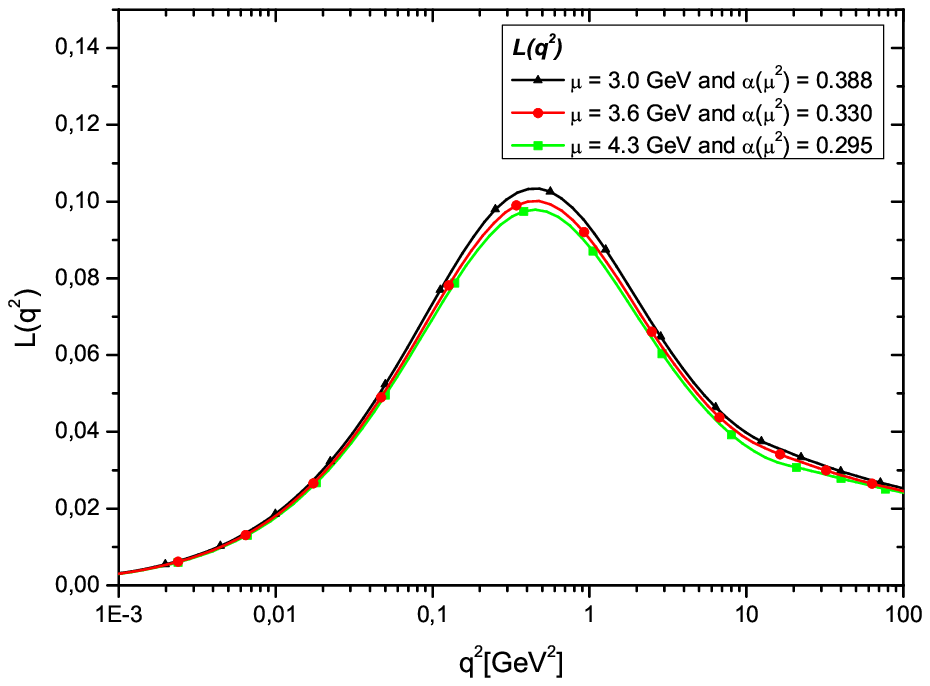}
\end{minipage}
\caption{{\it Left panel}: $-u(q^2)$ determined from Eq.~(24) at different renormalization points $\mu=3.0, 3.6, 4.3$ GeV. {\it Right panel}:  Same as in the previous panel but this time for $L(q^2)$.}
\label{fig3}
\end{figure}

Now, as discussed in detail in~\cite{Aguilar:2009nf},  
these (unrenormalized) equations must be properly renormalized, {\it i.e.},  
in such a way as to preserve the validity  of the BRST  identity in (\ref{BRST-1}), which should not be deformed by the renormalization process.
Note in fact that Eq.~(\ref{BRST-1}) constrains the cutoff-dependence 
of the unrenormalized quantities involved\footnote{ 
It is easy to recognize, for example,  by substituting into (\ref{simple}) 
tree-level expressions,
that  $F^{-1}(q^2)$ and  $u(q^2)$ have the same 
leading dependence on the ultraviolet cutoff $\Lambda$, namely
$F^{-1}_{\s{\mathrm{UV}}} (q^2) = u_{\s{\mathrm{UV}}} (q^2) = 
9 g^2/64 \pi^2 \log(\Lambda^2/q^2)$, 
while $L(q^2)$ is finite (independent of $\Lambda$).}.
Specifically, if we denote by $Z_c$ the ghost wave-function renormalization constant, with $Z_cF^{-1}_0=F^{-1}$ and with $Z_\Lambda$ the (yet unspecified) renormalization constant of the function $\Lambda_{\mu\nu}(q)$, with\footnote{The inclusion of the  $g_{\mu\nu}$ term is absolutely 
essential for the self-consistency of the entire renormalization procedure. To be sure, the 
$g_{\mu\nu}$ term appears naturally, given, for example, the form of the gluon propagator background quantum identity~\cite{Aguilar:2009pp}.} $Z_\Lambda[g^{\mu\nu}+\Lambda^{\mu\nu}_0]=g^{\mu\nu}+\Lambda^{\mu\nu}$, one finds that in order to preserve the identity (\ref{BRST-1}) one has to impose $Z_\Lambda=Z_c$~\cite{Aguilar:2009nf};  as a result, one finds
the relation
\be 
Z_c(\Lambda^2, \mu^2)[1+u_0(q^2,\Lambda^2)+L_0(q^2,\Lambda^2)] = 1+u(q^2,\mu^2)+ L(q^2,\mu^2).
\label{Zcren}
\ee
Imposing then the renormalization condition $F(\mu^2)=1$, 
going to Euclidean space, setting $q^2=x$, $k^2=y$ and $\alpha_s=g^2/4\pi$, 
and implementing the standard angular approximation,  
one finds the renormalized equations
\bea
F^{-1}(x) &=& Z_c - \frac{\alpha_s C_{\rm {A}}}{16\pi} \left[
\frac{F(x)}{x}\int_{0}^{x}\!\!\! dy\  y \left(3 - \frac{y}{x}\right) \Delta(y) 
+ \int_{x}^{\infty}\!\!\! dy \left(3 - \frac{x}{y}\right)\Delta(y) F(y) 
\right],\nonumber \\
1+u(x) &=&  Z_c - \frac{\alpha_s C_{\rm {A}}}{16\pi}\left[
\frac{F(x)}{x}\int_{0}^{x}\!\!\! dy\  y \left(3 + \frac{y}{3x}\right) \Delta(y) 
+ \int_{x}^{\infty}\!\!\! dy \left(3 + \frac{x}{3y}\right)\Delta(y)F(y) 
\right],
\nonumber\\
L(x) &=&  \frac{\alpha_s C_{\rm {A}}}{12\pi} \left[
\frac{F(x)}{x^2}\int_{0}^{x}\!\!\! dy\ y^2 \Delta(y) 
+ x \int_{x}^{\infty}\!\!\! dy \frac{\Delta(y) F(y)}{y}
\right].
\label{FGL}
\eea
From this last equation it is easy to see ({\it e.g.}, 
by means of the change of variables $y = zx$) that if $\Delta$ and $F$ are IR finite, then $L(0) = 0$, 
as mentioned before.

At this point one can substitute into the equations above the available lattice data on the gluon and ghost propagators thus determining in an indirect way the functions $u$ and $L$. The results are shown in Fig.~\ref{fig3}, from which we explicitly see the $\mu$-dependence of the KO function and in particular of the KO parameter, as well as the vanishing in the deep IR of the $L(q^2)$ form factor. Notice that the KO function saturates in the deep IR at the value\footnote{A value for the KO parameter of $-2/3$ has been in fact predicted in~\cite{Kondo:2009ug} by studying how the presence of the Gribov horizon affects the KO criterion. However since there is a residual dependence on $\mu^2$ of the KO parameter we consider this a coincidence due to the choice of the renormalization point rather than a fundamental prediction of the theory.} $u(0)\sim-0.6$.

\begin{figure}[!t]
\begin{center}
\includegraphics[scale=0.8]{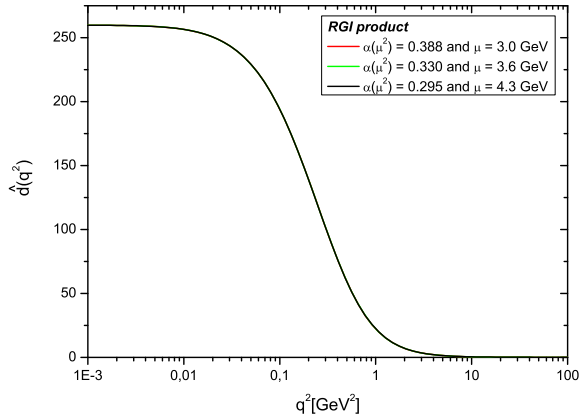}
\end{center}
\vspace{-.5cm}
\caption{The renormalization-group invariant product $\widehat{d}(q^2)$ obtained  combining the lattice results for the gluon propagator and our solutions for the function $u(q^2)=G(q^2)$ according to Eq.~(25)}
\label{fig4}
\end{figure}

Now, it is well-known that the product $g^2\widehat{\Delta}(q^2)$ is a RG-invariant quantity, representing the non-Abelian generalization of the QED quantity $e^2\Delta(q^2)$. Then using Eq.~(\ref{gBQI}) we find that the product
\be
\widehat{d}(q^2)=\frac{g^2(\mu^2)\Delta(q^2,\mu^2)}{[1+u(q^2,\mu^2)]^2},
\label{rgi}
\ee
is RG-invariant. But this then shows definitively that ({\it i}) the KO function cannot be an RG-invariant combination since ({\it ii}) its $\mu$ dependence must be such that it cancels the $\mu$-dependence of the numerator, as it is explicitly shown in Fig.~\ref{fig4}.

\begin{figure}[!t]
\begin{minipage}[b]{0.5\linewidth}
\centering
\includegraphics[scale=0.75]{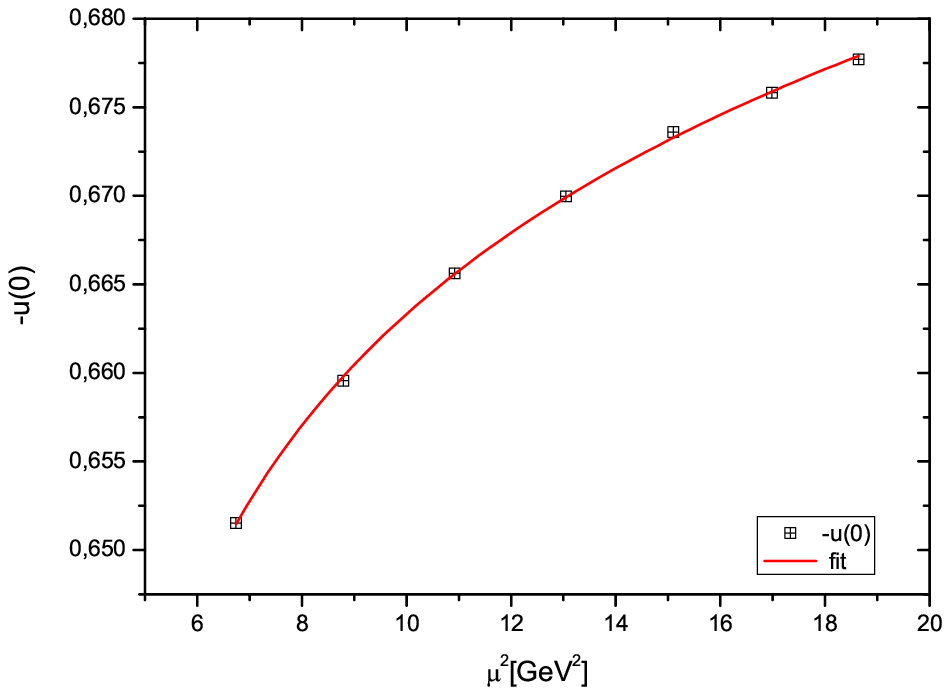}
\end{minipage}
\hspace{.5cm}\begin{minipage}[b]{0.45\linewidth}
\centering
\includegraphics[scale=0.75]{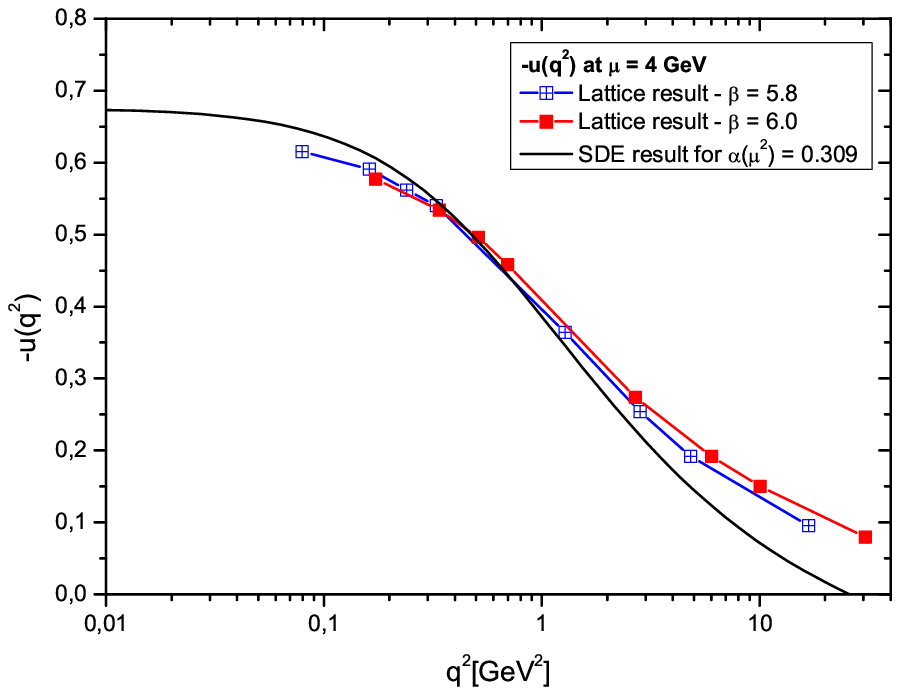}
\end{minipage}
\caption{{\it Left panel}: The dependence of the KO parameter $u$ on the renormalization
point $\mu$; the red solid line corresponds to a fit of a phase transition type $-u=a(\mu^2-b)^c$, with parameters  $a=0.633$, $b=3.57$, $c=0.025$.  {\it Right panel}:  The KO function, $-u(q^2)$, obtained from the solution of Eq.~(24) 
(solid black line) compared to the lattice data of ~\cite{Sternbeck:2006rd} at $\mu = 4\,\mbox{GeV}$}
\label{fig5}
\end{figure}

The $\mu$-dependence of the KO parameter is displayed in the left panel of Fig.~\ref{fig5}; the observed 
$\mu$-dependence is really sizable when contrasted with the impressive absence of any such 
dependence displayed by the genuinely RG-invariant quantity given in Eq.~(\ref{rgi}) which was 
computed using exactly the same sets of lattice data (Fig.~\ref{fig4}).
On the right panel of the same figure we show finally a comparison between our indirect determination of the KO function and the direct one  obtained in~\cite{Sternbeck:2006rd} where the function (\ref{id-1}) was studied by means of Monte Carlo averages: evidently the two curves compare rather well.

\begin{figure}[!b]
\begin{center}
\includegraphics[scale=0.8]{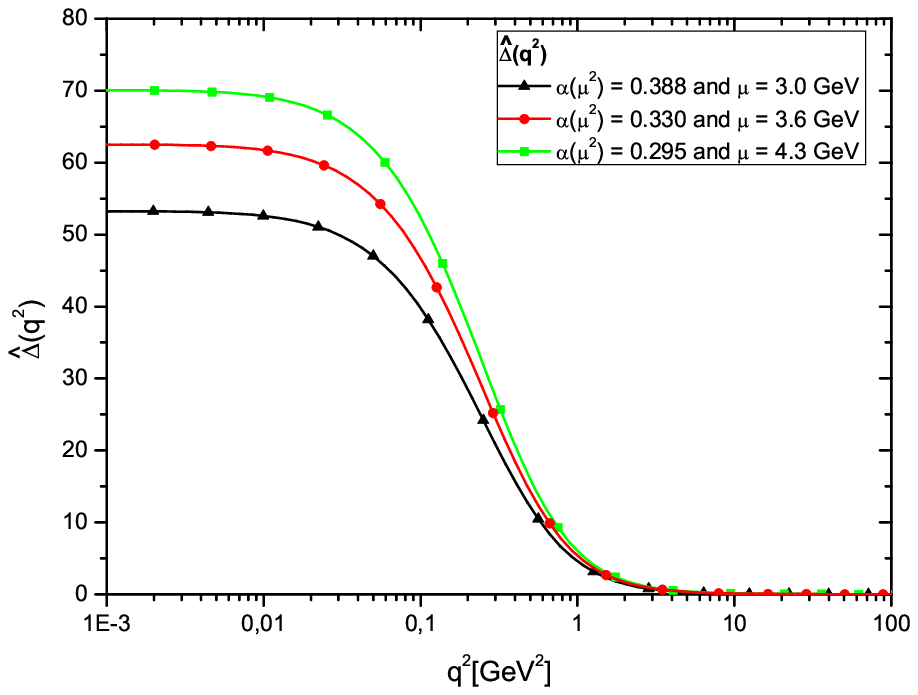}
\end{center}
\vspace{-1.3cm}
\caption{ The gluon propagator $\widehat{\Delta}(q^2)$ of the BFM, renormalized at
three different points: \mbox{$\mu = 3.0 \,\mbox{GeV}$} (black curve), \mbox{$\mu = 3.6 \,\mbox{GeV}$} (red curve) and \mbox{$\mu = 4.3 \,\mbox{GeV}$ (green curve)}.}
\label{fig6}
\end{figure}

Let us finally take a closer look at the background-quantum identity~(\ref{gBQI});  due to the central equality of 
Eq.~(\ref{KO-G}), we have that in the Landau gauge
\be
u(q^2) =  \sqrt{\frac{\Delta(q^2)}{\widehat{\Delta}(q^2)}} -1 .
\label{ulatt2}
\ee
Interestingly enough, this simple formula expresses the KO function in terms 
of two gluon propagators calculated in the Landau gauge of two very distinct 
gauge-fixing schemes, with no direct reference to the ghost sector of the theory.
This observation opens up the possibility 
of deducing the structure of the KO function using an entirely different, and completely novel, approach.
Specifically, one may envisage a lattice simulation\footnote{The lattice formulation of the background field method has been presented in~\cite{Dashen:1980vm}; interestingly enough, it was carried out in the Feynman gauge, which is the 
privileged gauge from the point of view of the pinch technique~\cite{Binosi:2002ft}.} of 
$\widehat{\Delta}$; then, $u(q^2)$ may be obtained from (\ref{ulatt2}) by simply  
forming the ratio of the two gluon propagators. 
Given that $\Delta(0)$ is found to be finite on the lattice~\cite{Cucchieri:2007md,Bogolubsky:2007ud}, it is clear that,
in order for the standard KO criterion to 
be satisfied ({\it i.e.}, $u(0)=-1$), $\widehat{\Delta}$ must diverge in the IR.  
Needless to say, we consider such a scenario highly unlikely. 
What is far more likely to happen, in our opinion, is to find a perfectly finite and well-behaved 
$\widehat{\Delta}$, which in the deep IR  
will be about an order of magnitude larger than $\Delta(0)$, furnishing a value $u(0)\sim -0.6$, namely   
what we have found in our analysis. In fact, one may turn the argument around: 
combining the results of this article with the lattice data for $\Delta$~\cite{Cucchieri:2007md,Bogolubsky:2007ud}, 
one may use (\ref{ulatt2}) to predict the outcome of the lattice simulation for $\widehat{\Delta}$; 
our prediction for the case of $SU(3)$ is shown in Fig.~\ref{fig6}.

\end{document}